\documentclass[%
 aip,
 amsmath,amssymb,
 reprint,%
]{revtex4-2}

\usepackage{graphicx}
\usepackage{dcolumn}
\usepackage{bm}
\usepackage[utf8]{inputenc}
\usepackage[T1]{fontenc}
\usepackage{mathptmx}
\usepackage{etoolbox}
\usepackage{url}            
\usepackage{subfiles}
\usepackage{array}
\newcolumntype{P}[1]{>{\centering\arraybackslash}m{#1}}
\usepackage{lineno}
\usepackage{microtype}
\usepackage{amsmath}
\usepackage{amssymb}
\usepackage{mathtools}
\DeclareRobustCommand{\bigO}{%
  \text{\usefont{OMS}{cmsy}{m}{n}O}%
}

\makeatletter
\def\@email#1#2{%
 \endgroup
 \patchcmd{\titleblock@produce}
  {\frontmatter@RRAPformat}
  {\frontmatter@RRAPformat{\produce@RRAP{*#1\href{mailto:#2}{#2}}}\frontmatter@RRAPformat}
  {}{}
}%
\makeatother
\begin{document}

\preprint{AIP/123-QED}

\title{Extended particle absorber for efficient modeling of intense laser-solid interactions}
\author{Kyle G. Miller}
\email{kylemiller@physics.ucla.edu}
\affiliation{Department of Physics and Astronomy, University of California, Los Angeles, California 90095, USA}

\author{Joshua May}
\affiliation{Department of Physics and Astronomy, University of California, Los Angeles, California 90095, USA}

\author{Frederico Fiuza}
\affiliation{SLAC National Accelerator Laboratory, Menlo Park, CA 94025, USA}

\author{Warren B. Mori}
\affiliation{Department of Physics and Astronomy, University of California, Los Angeles, California 90095, USA}

\date{\today}

\begin{abstract}
An extended thermal particle boundary condition is devised to more efficiently and accurately model laser-plasma interactions in overdense plasmas.  Particle-in-cell simulations of such interactions require many particles per cell, and a large region of background plasma is often necessary to correctly mimic a semi-infinite plasma and avoid electron refluxing from a truncated plasma. For long-pulse lasers of many picoseconds, such constraints can become prohibitively expensive.  Here, an extended particle boundary condition (absorber) is designed that instantaneously stops and re-emits energetic particles streaming toward the simulation boundary over a defined region, allowing sufficient time and space for a suitably cool return current to develop in the background plasma.  Tunable parameters of the absorber are explained, and simulations using the absorber with a 3-ps laser are shown to accurately reproduce those of a causally separated boundary while requiring only 20\% the number of particles.
\end{abstract}

\maketitle

\section{Introduction} \label{sec:intro}

Particle-in-cell (PIC) simulations have long been used to study the kinetic effects of laser-plasma interactions with overdense plasmas, with applications including the study of novel X-ray light sources,\cite{Watts2002DynamicsSpectra,Tsakiris2006RoutePulses,Park2006High-energyLasers,Ji2014EnergyInteraction,Capdessus2014Regime} generation of mono-energetic ion beams,\cite{Robinson2008RadiationPulses} experiments of collisionless shocks,\cite{Spitkovsky2008OnPlasmas,Martins2009IONSHOCKS,Fiuza2012Weibel-instability-mediatedLasers} transport experiments through warm-dense matter\cite{May2011MechanismInterface} and the fast ignition concept for inertial confinement fusion.\cite{Tabak1994IgnitionLasers,Tonge2009ALasers}  The laser-plasma interactions are often simulated for times on the order of picoseconds (1000s of laser periods) and for distances on the order of hundreds of microns (100s of laser wavelengths and 1000s of collisionless skin depths).  The thickness of the simulated overdense plasma region varies depending on the physical setup as well as the acceleration mechanism being explored.  For targets thin enough that the entire target can be simulated, vacuum regions are often used on either side of the target, consistent with the experimental setup.\cite{Silva2004ProtonInteractions,Sentoku2003AnomalousPlasma,Pukhov2001Three-DimensionalLaser,Lasinski1999Particle-in-cellApplications}  For thicker targets that cannot be simulated in their entirety, the plasma may be extended to the simulation boundary,\cite{Kemp2012InteractionPlasma,Yang1995AbsorptionBremsstrahlung,Pukhov1997LaserTargets,Wilks1993SimulationsInteractions} where an absorbing or thermal particle boundary condition is used.

When the laser-plasma interaction at the front of the target leads to large quantities of energetic electrons, a large flux of particles will in turn be found leaving the rear simulation boundary.  Independent of the particle boundary condition, the exiting stream of energetic particles can be problematic: either sharply removing the current (absorbing boundary condition) or the sudden stopping and accumulation of charge (thermal boundary condition) leads to an electric field buildup at the boundary.  This strong electric field will generate a return current that is carried by a hot, rarefied electron population (nearly symmetric to the incident electrons) instead of the proper cold, dense population.\cite{Bell1997Fast-electronExperiments}  The hot return current can both modify streaming instabilities that arise in the bulk plasma and modify the laser-plasma interactions at the front surface.   To avoid the electron refluxing, the plasma may be elongated such that the rear of the plasma is causally separated from the laser-plasma interaction region for some desired duration.\cite{Levy2013ConservationInteractions,Adam2006DispersionPlasmas}  In this case a small vacuum region is often placed to the right of the plasma to simplify the particle boundary conditions.  However, elongating the plasma introduces extra overhead from simulating the (often very particle-dense) excess material.

In an effort to preserve simulation integrity, while shortening the simulated plasma region, we propose an extended particle boundary condition that sporadically stops particles of certain energies over a defined distance.  In the presence of a low-density, hot particle beam or tail shooting into the plasma, this extended stopping avoids localized charge buildup or current deficiency that occurs when using an absorbing or thermal boundary condition, thus allowing a suitably cool return current to develop in the background plasma over an extended period of time and space.  The initial idea of this extended absorbing boundary condition was briefly described alongside some results in references,\cite{Tonge2009ALasers,Kemp2014LaserplasmaIgnition} but here we provide details for implementation, improvements, potential issues and best practices of such an absorber.

The outline of this paper is as follows.  We first discuss in Sec.~\ref{sec:overdense} the possible issues with truncating a semi-infinite plasma with standard reflecting, absorbing, or thermal bath particle boundary conditions. In Section~\ref{sec:absorber} we present the concept and design of the absorber, along with the parameters that can be specified.  Finally, in Sec.~\ref{sec:results} we present simulation results from the PIC code \textsc{Osiris}, where we test the implementation of the absorber boundary condition on a finite target against a semi-infinite (causally separated) plasma.

\section{Boundary issues in overdense plasma simulations} \label{sec:overdense}

The motivation for this work is to efficiently model the interaction of a high-intensity laser with the surface of an overdense plasma. This interaction generates copious amounts of relativistic electrons that propagate forward deep into the target. The forward-going electrons lead to a return current of electrons that then interacts with the laser.  When investigating how a laser is absorbed into relativistic electrons, ideally only the region of interest need be simulated; such a region might be an underdense or vacuum region in front of the target or a location some distance into the target itself.  However, this presumes that the spectrum of electrons---including its currents and heat flux---are the same as if the entire plasma region was simulated.  This may not be the case if the boundary conditions at the edge of the simulation do not properly represent the actual conditions.

The preference of only modeling a small region of a larger plasma is more important for higher target densities, which require smaller time steps and cell sizes (compared to the laser period and wavelength) due to increased plasma frequency and decreased scales of physical interest.  Even in simulations that include a larger transport region for the electrons, it may still not be feasible to simulate the entirety of a physical target---for instance, a millimeter-scale solid-density target. For such simulations to be reliable, it is of course necessary that they not be affected by the choice of the simulation boundary location and the associated boundary conditions. Therefore, it is critical to find a boundary condition that mimics the effects of a quiescent plasma of unbounded depth in the direction opposite from laser incidence.  Such a boundary condition is useful even for targets not fully described by an infinite thickness, in that the physics may be understood at least to some approximation as a superposition of unbounded and finite-thickness effects.

Depending on the laser pulse length and/or the target size, it is sometimes possible to simulate an infinitely deep quiescent plasma by expanding the simulated plasma to distances causally separated from the interaction---or perhaps half that far, so signals (moving near the speed of light) cannot reach the boundary and return.  Such simulations allow us to determine the ``correct'' physics, against which we will compare our results.  However, in practice these simulations are usually impractical.
The required plasma thickness also scales with pulse length, so simulation computer time scales quadratically with the pulse length for one- and two-dimensional periodic cases. For finite-width cases, the transverse size may also need to be extended depending on the pulse length.
Therefore, a compact target that reproduces the behavior of a larger or infinite target is desirable.

To make clear the need for an appropriate boundary condition, we mention two spurious effects that can occur in simulations of a truncated target. First, we observe that the laser-matter interaction continuously deposits energy into the plasma. Although the details may be complex, we can assume that the energy will somehow diffuse or dissipate deep into the target; if this is not possible (e.g., inhibited by a boundary), the target will heat artificially. As the absorption of the laser has been postulated to be highly dependent on the target temperature in some scenarios,\cite{May2011MechanismInterface} this heating can feed back into the absorption itself and greatly affect the overall simulation.

The second effect of an improper boundary is the modification of the plasma distribution function long before any heating of the bulk electrons through an effect known as electron refluxing.\cite{Sentoku2003HighTarget,Quinn2011RefluxingPulses}  Electrons accelerated by a laser generally make up a super-thermal tail in the distribution function, extending to energies far greater than the background thermal energy.\cite{Wilks1992AbsorptionPulses} These high-energy electrons have low collisionality (even in solid-density targets\cite{Nilson2009BulkInteractions}) and may thus travel nearly ballistically through the material in the absence of strong fields. At a given transverse plane in the plasma, the bulk plasma exhibits a small backward drift to provide a return current. However, at the plasma boundary the return current formation can be more complicated. For finite-thickness targets with a vacuum region on the far side (both in the lab and in simulations), the first electrons to reach the back of the material exit into a vacuum and continue along a ballistic path. However, an electrostatic field can quickly build at the target rear surface; no such field grows inside the target due to the high conductivity of the background plasma.  This decelerating field grows in time, eventually causing energetic electrons to be reflected back into the target.  We refer to the re-injection of electrons from this electrostatic field as electron refluxing.  After this reflection, the electrons again travel ballistically though the target, where they can reach another vacuum boundary region (sides) and go through another reflection; alternatively they can re-enter the laser-plasma interaction region, in which case their large kinetic energy modifies how they interact with the laser, perhaps significantly.

The reflecting electrostatic field responsible for electron refluxing in finite-thickness targets arises due to the adjacent vacuum region, where escaping electrons leave behind a net charge that resides on the surface. We have found that standard PIC particle boundary conditions responding to high-energy particle beams actually exhibit reflectivity similar to a vacuum boundary.  In particular, so-called ``reflecting,'' ``absorbing,'' and ``thermal re-emitting'' boundaries all lead to refluxing particles early in the beam interaction; in each case a strong electric field builds up at the boundary.  A specular reflecting boundary condition---where the sign of the momentum perpendicular to the boundary wall is reversed---clearly leads to refluxing; however, no electrostatic field is developed at the boundary since the boundary itself reflects the particles.

An absorbing particle boundary, in somewhat simplified terms, simply removes exiting particles (and their corresponding current) from the simulation space.  However, electromagnetic PIC codes like \textsc{Osiris} advance the fields forward in time via Faraday's and Ampere's laws while depositing current such that the continuity equation is rigorously satisfied. Therefore, when a particle's current disappears, its charge is in fact frozen at its last location.  In the case of an absorbing boundary, an exiting electron beam will cause a static charge buildup.
The bulk of the plasma near the boundary attempts to shield out the boundary field within a few Debye lengths.  However, the field continues to build as more current crosses the boundary and can eventually become large enough to accelerate background electrons backward at relativistic energies, leading to a hot tail of refluxing electrons.
Vacuum boundaries behave similarly, where ions are slowly driven off the target via what is known as target normal sheath acceleration\cite{Wilks2001EnergeticInteractions,Mora2003PlasmaVacuum} (TNSA); again, electrons reach the rear edge in greater numbers than the ions which are able to escape.

A thermal bath particle boundary---where particles are re-emitted into the simulation space with momentum sampled from a specified thermal distribution---may at the outset seem to remedy the issues caused by reflecting and absorbing boundary conditions. We find, however, that a thermal boundary fails to reduce the artificial refluxing. First of all, the correct bath temperature is somewhat ambiguous, and we observe that an incorrect choice leads to clearly incorrect behavior. A bath that is too hot will artificially heat the background electrons in the target; one that is too cold re-emits the particles with too little thermal velocity to diffuse back into the box, manifesting errors similar to those of an absorbing boundary. A drifting Maxwellian moving back into the box (attempting to maintain current neutrality) could be used in place of the stationary Maxwellian generated by the thermal bath. However, determining the proper drift and thermal velocity of this modified Maxwellian is difficult. Furthermore, if a ``correct'' bath temperature and drift velocity were known from a causally separated simulation, it would be a function of time.  Although there are ways to measure effective temperatures (as discussed in Appendix~\ref{app:temp}), the distribution function is not necessarily well-described by single temperatures; thus it is unclear what one should measure nor how to specify the bath temperature.

The extended boundary condition we describe in this paper is designed with the goal of self-consistently generating a neutralizing, drifting background distribution that imitates an infinite plasma as closely as possible.

\section{Absorbing boundary region} \label{sec:absorber}

We desire to model a semi-infinite target with a finite simulation.  We must therefore find a way to remove the energetic beam electrons from the simulation while providing a self-consistent return current.  One way to do this is to stop the relativistic electrons not at a single point (as in the absorbing boundary condition), but over an extended region.  The electrostatic field is then dispersed over a sufficient ``volume'' such that the electric field driving the reflux is spread out and reduced in amplitude. Thus, a larger ``volume'' of bulk plasma is driven backwards without reflecting the high-energy electrons.  In essence, we create an extended boundary condition for the particles.  The plasma can then shield out the field more rapidly such that the potential does not build up over time.

To create this extended boundary condition, we denote some region near the back of the simulation box to be the absorbing region.\footnote
{
 The absorbing region can also be a finite region in the middle of the plasma.  This was done in Ref.~\onlinecite{Tonge2009ALasers} for integrated fast ignition simulations.  In this case the electrons were slowed gradually by a drag force.
}
Within this volume, high-energy particles are selected at random and instantaneously thermalized to a given temperature in a single time step.  We do this rather than slowing down the particles gradually---arguably more physically correct---because it allows us to stop the particles throughout the absorber volume without any knowledge of the global beam characteristics; it also obviates the need to track particles across multiple distributed-memory processes as they decelerate (more algorithmically complex).  The disadvantage of stopping the particles instantaneously is the emission of Bremsstrahlung radiation.\cite{Allen1973AstrophysicalQuantities}  However, this radiation is in most cases both poorly resolved on the simulation grid and can effectively be eliminated by higher-order particle shapes and smoothing.

We define a stopping loop as the process of iterating over all particles in the defined absorbing region, calculating the probability that each particle will be stopped, then stopping a particle (i.e., directly changing its momentum) if a randomly generated number is less than that probability.  Stopping loops are performed at a defined time interval, $t_a$ (can be one or multiple simulation time steps).  At a given position, the probability of particle stopping and corresponding re-emission are controlled by three parameters: an energy threshold, a mean free path, $\lambda$, and a re-emission temperature.  The energy threshold provides a way to distinguish between background particles (energy is below the threshold, nothing happens) and hot electrons (energy is above the threshold, stopping may occur).  The mean free path, particle velocity and time between stopping loops, $t_a$, are used to calculate the probability of stopping.  This should be done such that the mean distance traveled by an ensemble of particles before stopping is given by the mean free path, but that no energetic particles reach the particle boundary.  The re-emission temperature is used to define the Maxwellian distribution from which the particle will be re-emitted in each of the three dimensions.  For both the energy threshold and the re-emission temperature, we allow the user to specify either absolute numbers or to calculate the parameters as multiples of the local plasma temperature near a given particle.  If required, the local plasma temperature is calculated by integrating the absolute value of the proper velocity in each dimension over all particles in a given cell.  See Appendix~\ref{app:temp} for further details on how this is done.  A typical value for the energy threshold is six times the local thermal velocity in any direction, and the re-emission temperature is usually just set to be equal to the local thermal velocity in every direction.  Below we discuss two possible methods to accomplish the desired stopping.

\subsection{Hazard function stopping} \label{sec:hazard}

\begin{figure}
\includegraphics[width=\linewidth]{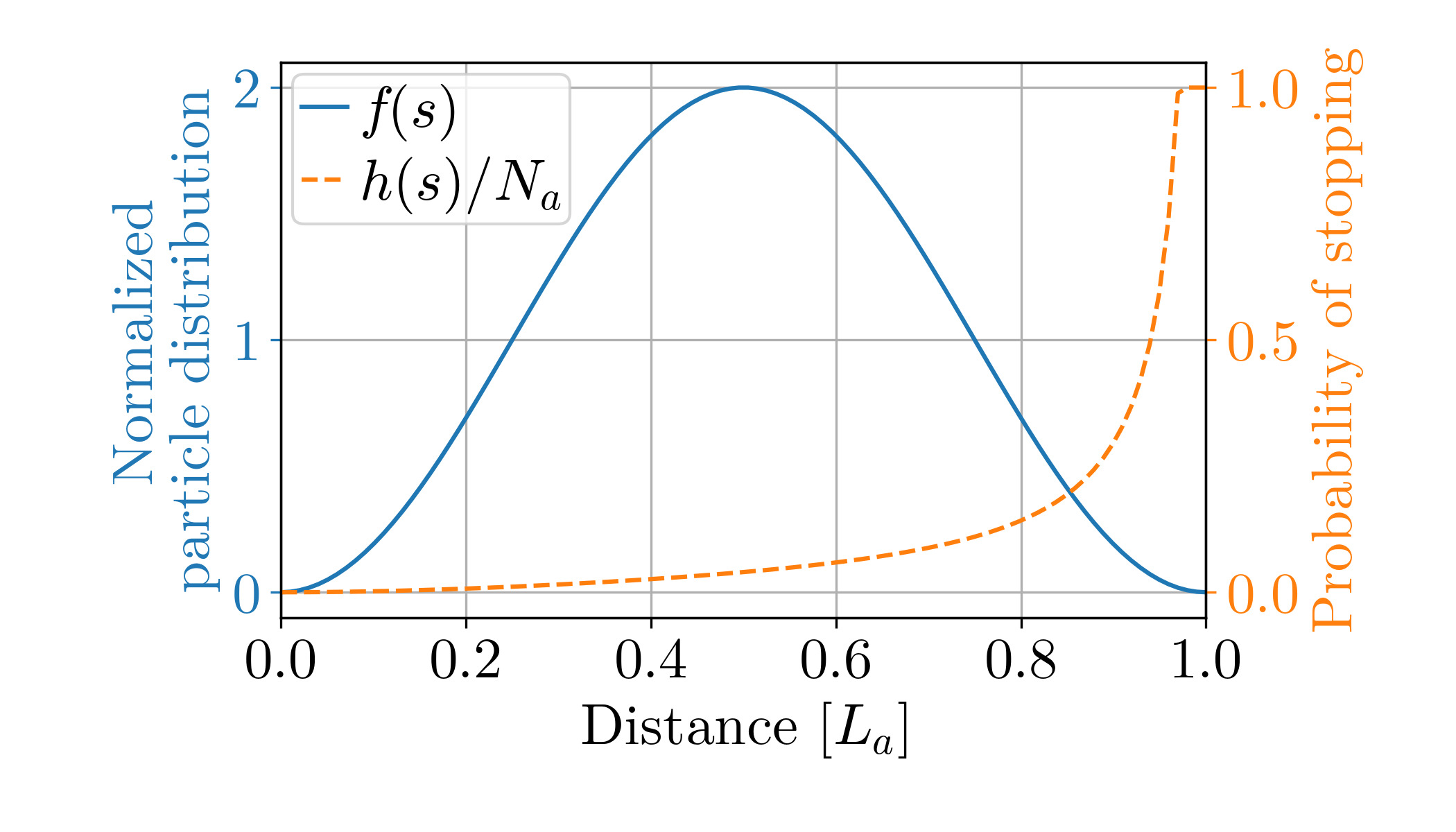}
\caption{\label{fig:f-and-h} Normalized probability density function (blue) for desired particle distribution, along with the probability of stopping (orange, dashed) for $N_a=100$ stopping loops over the absorbing region. Here the mean free path is $\lambda=0.5\,L_a$.}
\end{figure}

Let us assume that all the particles we wish to stop are streaming near the speed of light, $c$, in the positive $x$-direction.  Rather than stopping these particles at a single boundary that is one cell thick (like a traditional thermal boundary condition), we aim to stop these particles one at a time over the length of an absorbing region many cells thick.  Thus particles are stopped over time and space.  To facilitate this, we can define a desired normalized distribution that specifies the density of particles in space once they are all stopped.  This is in fact a probability density function, call it $f(s)$.  For example, we could use the function
\begin{equation} \label{Eq:prob}
    f(s) = 2\sin^2(\pi s),
\end{equation}
where $s \equiv x/L_a$ for a particle that has traveled distance $x$ into the absorbing region of length $L_a$.  Defined in this way, all particles will have stopped after a distance of $L_a$.  We can also consider a parameterization in time as $s \equiv t/T_a$ for a particle that takes a time $T_a$ to traverse the absorbing region (i.e., $T_a=L_a/v_x$ for a particle velocity $v_x$ in the direction of the absorbing region).  The associated cumulative distribution function is then
\begin{equation}
    F(s) = \int_0^s f(s')\,ds' = s - \frac{\sin ( 2 \pi s )}{2\pi},
\end{equation}
and the (continuous) probability that a particle will stop in this interval is given by the hazard function $h(s)$, defined as\cite{Evans2000TermsSymbols}
\begin{equation} \label{Eq:hazard}
    h(s) \equiv \frac{f(s)}{1-F(s)} = \frac{4\pi \sin^2 ( \pi s )}{2\pi ( 1-s ) + \sin ( 2 \pi s )}.
\end{equation}
Since we stop particles instantaneously at discrete intervals rather than gradually slowing them over some distance, we assume that a given particle will have been selected to be stopped after at most $N_a=L_a/t_av_x=T_a/t_a$ stopping loops (within each stopping loop only a small number of selected particles are stopped). Thus the probability of stopping for each hot particle located at position $s$ for the hazard function absorber is
\begin{equation} \label{Eq:P-haz}
    P_\mathrm{haz}(s) = \begin{dcases} 
      h(s)/N_a = h(s)t_a/T_a & s < 1 \\
      1 & s \geq 1
   \end{dcases}.
\end{equation}
The probability density function given in Eq.~(\ref{Eq:prob}) is shown along with the probability of stopping for $N_a=100$ in Fig.~\ref{fig:f-and-h}.  Note that the mean free path for this case is $\lambda=L_a/2$, and thus the scheme can also be parameterized by $s=x/2\lambda$.

Stopping via a hazard function works well when the hot particles are all streaming in the same direction and have large velocity components mainly in one direction (i.e., $|\mathbf{v}| \approx v_x$).
However, many simulations employing the absorber may exhibit forward-going particle beams with non-negligible transverse momentum.  Since the hazard function absorber stops particles based only on their longitudinal momentum, the background plasma can become disproportionately and unstably hot in the transverse direction.  If transversely hot particles have very low longitudinal momentum or are located near the front of the absorbing region, they will have a very low probability of stopping and can continue to drift unchecked for long periods of time (especially dangerous for transversely periodic simulations).  In our test simulations the growing transverse currents could eventually form a large reflecting magnetic field near the beginning of the absorbing region, resulting in a reflux of the hot particle beam.  To prevent any unstable transverse momentum growth, we propose the following alternative algorithm that stops hot particles without assuming a specific velocity distribution (e.g., all moving forward near the speed of light).


\subsection{Linearly varying absorber} \label{sec:linear}

A much simpler approach is to stop particles based on their energy alone, independent of their direction of motion or position in the absorbing region.  Let us assume that most of the energetic particles we wish to stop are traveling mainly in the forward $x$-direction at speed $c$ (though some may be travelling backward or have large transverse velocities).  Suppose that, similar to Sec.~\ref{sec:hazard}, our absorbing region begins a distance $L_a$ from the right simulation edge and that we wish to have a mean free path of $\lambda<L_a/2$ (now in any direction) for the particles, such that they are all stopped before the simulation boundary.  An intuitive way to accomplish this is to perform a large number of stopping loops, $N_a \approx 100$, in the time it takes an energetic particle to travel a distance of $x_f$ through the absorbing region; we can naively set the stopping probability to $P = 1/N_a$ for each particle in each loop.  We thus require the minimum time between stopping loops to be $t_a = x_f/N_a c$.
Such a scheme actually corresponds to a hazard function of $h(x)=1/x_f$ and a probability density function of $f(x)=\frac{1}{x_f} \exp \big( -\frac{x}{x_f} \big)$.  However, after $N_a$ stopping loops (i.e., after particles propagate a maximum distance of $N_a c t_a = x_f$), integrating $\int_0^{x_f} f(x)\,dx$ shows that on average only 63\% of particles will have been stopped using the $P=1/N_a$ probability.  We would then need to use the rest of the absorbing region to stop the remaining hot particles.  We can extend this idea to construct a more effective absorber that meets our needs.

Based on the above argument, the stopping probability for a particle with velocity magnitude $v=|\mathbf{v}|$ would be $P(v) = vt_a/x_f$.  Note that the velocity used is independent of direction, providing equal stopping for particles traveling rapidly forward, backward, or transversely to the absorbing region.  However, to facilitate the stopping of particles past a distance of $x_f$, we parameterize the absorbing region by two longitudinal positions: (1) a location specifying the start of the absorbing region (defined as $x=0$ here) and (2) a location specifying where the absorber is at its full strength, defined as $x_f$.  Variables such as the mean free path, energy threshold and re-emission temperature are defined at positions (1) and (2); we will refer to these two kinds of parameters as variables of type ``start'' and type ``full'' ($x_f$ refers to the full position).  In front of the start position, the absorber is turned off.  In-between the start and full positions, the stopping parameters (e.g., energy threshold and mean free path) are changed linearly from their start values to their full values.  From the full position to the back of the box, the stopping parameters remain at their full values.  The absorber is designed to start with modest stopping parameters and increase to more stringent stopping in order to avoid any sharp transitions in the simulation physics that may result in wave reflections or other spurious effects.  Employing a linear ramp comes with the added benefit that we can use a similar set of input parameters for wide range of simulations with varying beam characteristics.

To be specific, we define a scale length, $L(x)$, that varies linearly between the start and full scale lengths, $L_s$ and $L_f$, respectively, over the distance $x_f$:
\begin{equation} \label{Eq:length}
    L(x) = \begin{dcases} 
      L_s + 2(L_f-L_s)x & x < x_f \\
      L_f & x \geq x_f
   \end{dcases}.
\end{equation}
The probability of stopping for this linearly varying absorber is then
\begin{equation}
    P_\mathrm{lin}(x,v) = vt_a/L(x).
\end{equation}
We then let $L_s=x_f$ and define $L_f$ such that particles are rapidly stopped, e.g., $L_f=(L_a-x_f)/20$.


Using this method, instead of initially specifying the probability density function $f(s)$ and then finding the hazard function $h(s)$, we are specifying $h(s)$ first.  One can solve for $f(s)$ using Eq.~(\ref{Eq:hazard}) and its derivative in $s$, assuming a piecewise form for $h(s)$.  First we write down the hazard function in normalized coordinates as
\begin{equation} \label{Eq:haz_lin}
    h(s) = \begin{dcases} 
      \frac{1}{\lambda_s + 2(\lambda_f-\lambda_s)s} & s < s_f \\
      \frac{1}{\lambda_f} & s \geq s_f
   \end{dcases},
\end{equation}
where similarly to Sec.~\ref{sec:hazard} we have $s\equiv x/L_a$, $\lambda_i = L_i/L_a$ and $s_f=x_f/L_a$.  To solve for the probability density function, we require $f(s)$ to be a continuous piecewise function and to integrate to 1, yielding the result
\begin{equation} \label{Eq:f_lin}
f(s) = \left\{ \def\arraystretch{1.4} \begin{array}{@{}l@{\quad}l@{}}
      s_f \cdot \left( \lambda_s s_f \right)^{\frac{s_f}{\lambda_f-\lambda_s}} \cdot & \\
      \quad \left( \left[ \lambda_f-\lambda_s \right]s + \lambda_s s_f \right)^{\frac{s_f}{\lambda_s-\lambda_f} - 1} & \smash{\raisebox{.7\normalbaselineskip}{$s<s_f$}}  \\
      \frac{1}{\lambda_f} \cdot \exp{\frac{s_f-s}{\lambda_f}} \cdot \left(\frac{\lambda_f}{\lambda_s}\right)^{\frac{s_f}{\lambda_s-\lambda_f}} & s\geq s_f \\
    \end{array} \right. .
\end{equation}

\begin{figure}
\includegraphics[width=\linewidth]{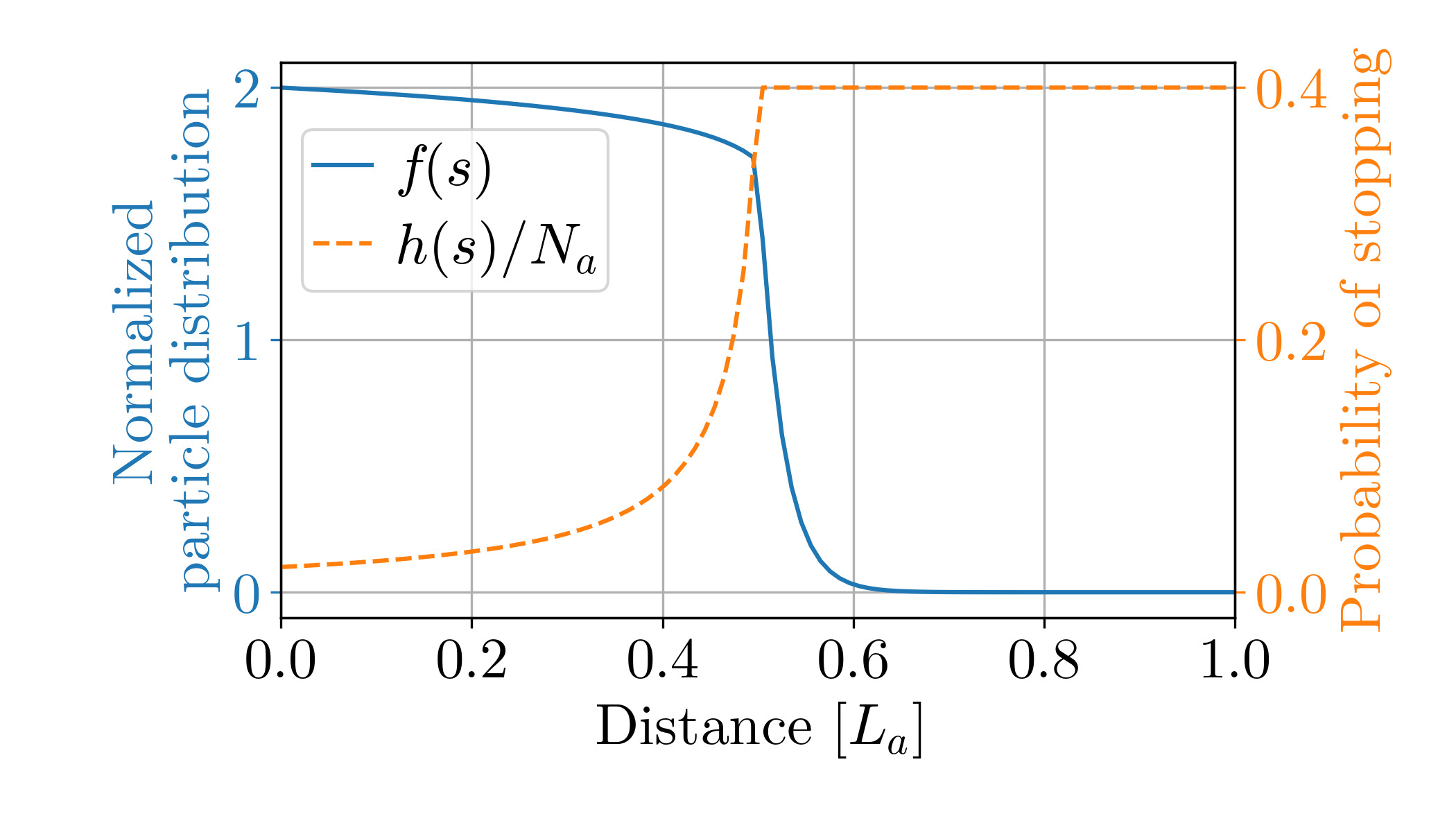}
\caption{\label{fig:f-and-h-lin} Normalized probability density function (blue) and probability of stopping (orange, dashed) from Eqs.~(\ref{Eq:haz_lin}) and (\ref{Eq:f_lin}) with $\lambda_s=0.5$, $\lambda_f=0.025$ and $s_f=0.5$. Here we have 200 stopping loops over the entire interval, and the mean free path is $\lambda=0.257\,L_a$.}
\end{figure}



The probability density function and hazard function for this linearly varying absorber are plotted in Fig.~\ref{fig:f-and-h-lin} for $\lambda_s=0.5$, $\lambda_f=0.025$ and $s_f=0.5$.  
In contrast to the previous scheme, many particles are stopped right at the beginning of the absorbing region (small, non-zero probability of stopping but large number of particles), and the mean free path of the electrons is $\lambda=0.257\,L_a$ instead of $0.5\,L_a$ for this case.  The overall length of the absorbing region and individual parameters can be adjusted to give the desired profile, but care should be taken so that very few electrons ever reach the right boundary.  This linearly varying scheme is much more flexible and reliable than the previous scheme.  By stopping based on absolute velocity magnitude, particles are stopped whether traveling forward, backward or transversely to the absorbing region.  Specific results will be shown and discussed in Sec.~\ref{sec:results}.

\subsection{Appropriate mean-free-path length} \label{sec:concept-mfp}

We wish to estimate an appropriate mean free path to use with the absorbing boundary before running the simulation.  To do so we will first make an argument based solely on relevant simulation parameters, then make a second based on the strength of the stopping electric field.

First, typical two-dimensional simulations of this type have square cell sizes of length $0.2\,c/\omega_0$, where $\omega_0$ is the laser frequency, with an accompanying time step of $\Delta t \lesssim 0.141\, \omega_0^{-1}$.  For simulations such as ours the overdense plasma density is $n=10n_c$, where $n_c$ is the critical density.  If a minimum of 50 stopping loops is desired over a mean free path (100 loops over the absorber length), this sets a minimum value of $\lambda \approx 22\,c/\omega_p$.  However, performing the stopping routine at every time step can be computationally expensive; thus stopping every $m$ time steps would increase the minimum mean free path to $\lambda \approx 22m\,c/\omega_p$.

The physics arguments that set the scale of $\lambda$ can be understood as follows. Collisionless plasmas attempt to remain current neutral. If a net current exists at some instant in time and space, then electric fields are induced on the electron-frequency time scale to neutralize the current. However, this process is more complicated at a boundary. Consider an absorbing boundary; as electrons leave, a net charge builds up, generating a repelling electric field. There is no background charge outside the boundary that can move in to cancel this field. As a result the electric field accelerates background electrons just inside the boundary, growing until it can accelerate electrons backward at near the speed of light. A thermal bath boundary condition results in a very similar situation. Once an electron leaves, another is remitted from a specified distribution function. The probability is chosen based on balancing the flux of a thermal plasma. Therefore, if the electrons leaving come from a hot tail, the remitted electrons cannot properly cancel this current.  A large electric field and potential build up, generating a reflux of electrons with relativistic energies.  However, if we instead stop the electrons over a distance along which the plasma can naturally generate a return current, then the physics will most closely resemble reality. The thickness must then be several skin depths wide, i.e., $\lambda \gtrsim \bigO(10\,c/\omega_p)$, as a skin depth sets the scale over which the current from relativistic electrons is neutralized. This is on the same order of magnitude as what was estimated above for numerical reasons.

Both the numerical and physical arguments predict a mean free path of about the same order, and in Sec.~\ref{sec:results} we explore the performance of absorbers of various size.

\subsection{Particle splitting and recombination}

The absorber is required when simulating overdense laser-plasma interactions because there is a large flux of relativistic electrons moving into the plasma. For such situations there is another issue that must be addressed simultaneously. Due to the relatively large charge (they represent many real electrons) of macroparticles in PIC simulations, these hot particles can generate artificially large wakes, causing them to slow down as they propagate through the plasma.\cite{Tonge2009ALasers,May2014}  These wakes can also lead to larger levels of turbulence. To avoid the macroparticle-stopping issue, we also run with the particle splitting algorithm discussed in Ref.~\onlinecite{May2014}.  This involves defining an energy threshold above which energetic electrons will be split into smaller particles to avoid significant energy loss.  Split particles are given a small boost in momentum space (typically 1\%) so that they don't travel along identical trajectories.

However, we anticipate that these numerous small particles will eventually be stopped somewhere in the absorbing region.  This can lead to a load imbalance in which a disproportionately large number of particles are distributed over a small number of processors.  Since energy conservation is already slightly violated by the stopping of fast particles, this recombination is done simply by averaging the momentum of two particles at a time that lie in the same octant of momentum space.  We only look to recombine such particles in the region where particle stopping occurs, and only particles with a charge less than the charge of an unsplit particle are considered for recombination.

\section{\textsc{Osiris} simulation results} \label{sec:results}

To demonstrate the effectiveness of the absorber region, we present results from a variety of 2-dimensional simulations of a laser incident on an overdense plasma.  Simulations were done using \textsc{Osiris},\cite{Fonseca2002} where an absorbing region has been implemented.

\subsection{Simulation setup}
\begin{figure}
\includegraphics[width=\linewidth]{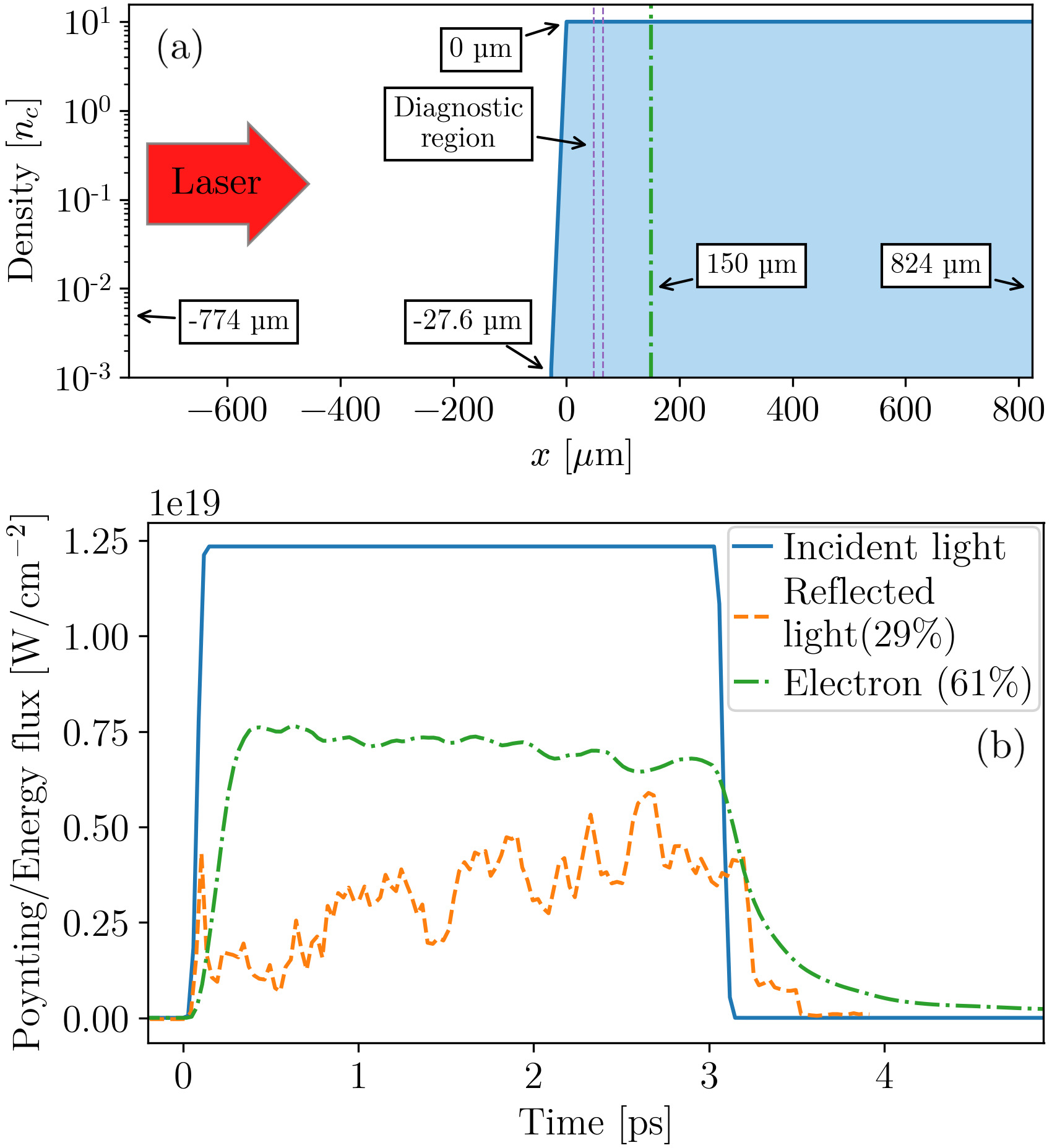}
\caption{\label{fig:laser} (a)~Simulation schematic, showing the full box size. The box is truncated at 150~$\mu$m when the absorber is in use. (b)~Laser Poynting flux incident at the plasma critical interface, reflected Poynting flux measured 387~$\mu$m to the left of the critical interface and forward electron energy flux measured over the diagnostic region.  All quantities are synced up in time for better visualization.  Percentages represent integrated energy flux as a fraction of the total incident energy.}
\end{figure}

In the simulations, an intense 1-$\mu$m plane-wave laser with normalized amplitude $a_0=3$ and 3~ps in duration (2.9-ps flat envelope with 0.13-ps rise and fall ramps) is incident on uniform plasma with density $n=10n_c$ (where $n_c$ is the critical density).  The exponential ramp has a scale length of 3~$\mu$m and begins at $x=-27.6$~$\mu$m.  The critical density is then located at $x_c=-6.9$~$\mu$m, and we define time $t=0$ to be when the leading edge of the laser pulse would arrive at $x_c$ if traveling at speed $c$. The laser is focused to the critical surface and is launched from the left wall.  The plasma skin depth is $c/\omega_p = 50.3$~nm and $c/\omega_0 = 159.2$~nm.  See Fig.~\ref{fig:laser}(a) for a schematic.

The simulations used periodic boundary conditions in the second dimension ($y$), and the laser was polarized with its electric field in the simulation plane (p-polarized). The simulation dimensions were kept constant in the $y$-direction,  3.2~$\mu$m, and in the $x$-direction were either 923.9 or 1597.8~$\mu$m for truncated and causally separated runs, respectively.  Square cells of size 0.2~$c/\omega_0$ were used, resulting in a simulation domain of 50197$\times$100 cells for the simulation with the largest length in $x$ (29025$\times$100 cells otherwise).  The time step was 0.141~$\omega_0^{-1}$.  The electron (ion) species had 64 (16) particles per cell, and each species used cubic interpolation with an initial temperature of 0.1~keV.  We employed a static load balancing routine\cite{Fonseca2013} at initialization to distribute processing elements in an optimal configuration, and the particle push time was delayed until the laser neared the plasma.

\begin{figure}
\includegraphics[width=\linewidth]{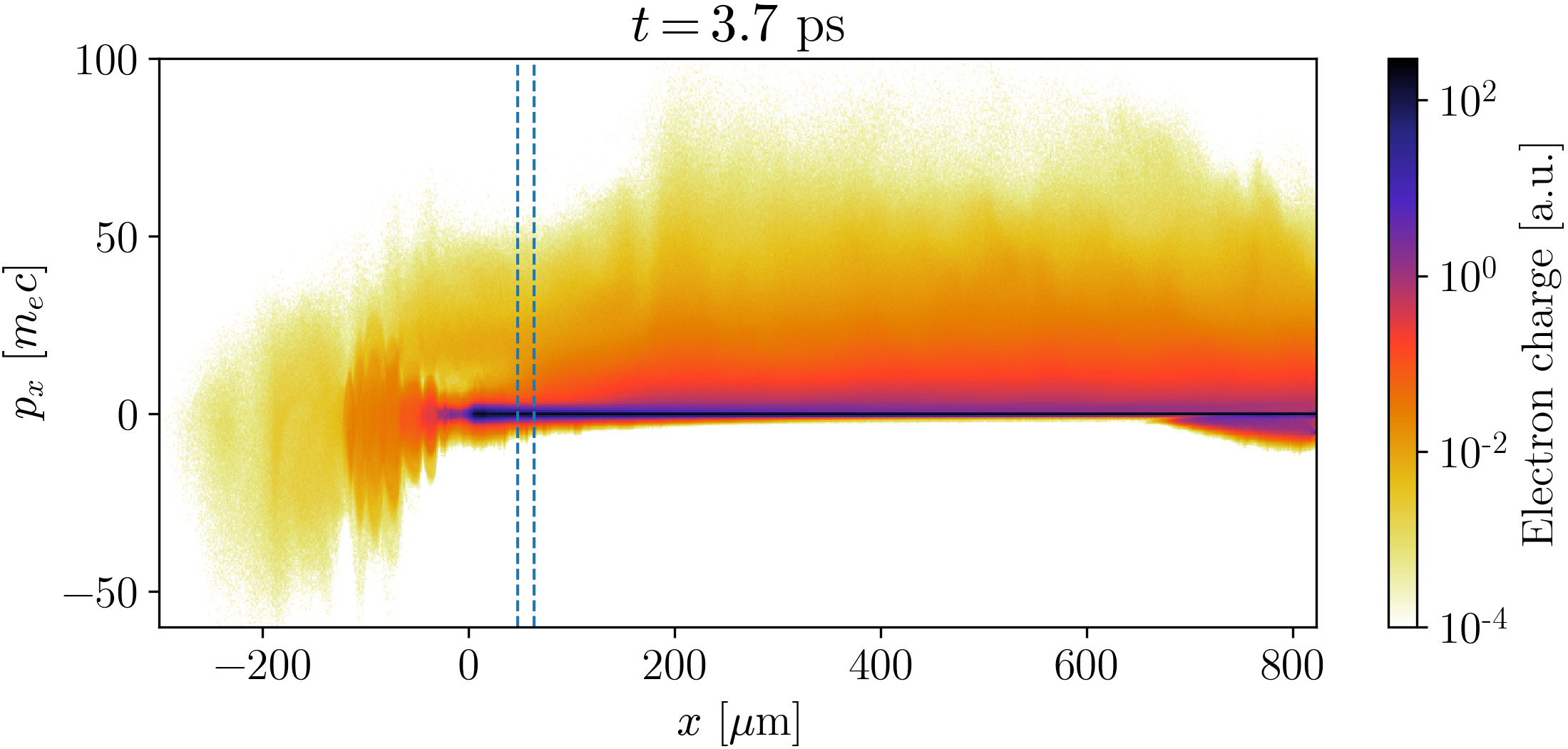}
\caption{\label{fig:px-x-a} The $p_x$-$x$ phasespace for the causally separated simulation (single run, not averaged).  The dashed lines indicate the diagnostic region, but the plasma has to be much larger in length to be causally separated from the hot return current reflecting off the right boundary.}
\end{figure}

In Fig.~\ref{fig:laser}(b) we show the temporal laser profile, as well as the reflected Poynting flux and
transmitted particle energy flux.  The reflected Poynting flux is calculated by measuring the total Poynting flux 380~$\mu$m before the critical-density interface, then subtracting the known incident laser flux.  Both the Poynting and energy fluxes plotted in Fig.~\ref{fig:laser}(b) are translated in time to line up with the incident laser light.  To diagnose the forward momentum and energy flux deep in the plasma, we choose a diagnostic region 48--64~$\mu$m into the uniform plasma over which we average the particle data in space.  The energy flux is defined as $\int (\gamma-1)m_ec^2 \mathbf{p}/\gamma\,d\mathbf{p}$ for electron mass $m_e$.  In order to avoid particle refluxing from either boundary in the $x$-direction, a 746-$\mu$m vacuum region (computationally inexpensive because of the static load balancing) is placed to the left of the plasma upramp, and the uniform-density plasma is extended to the right a distance of 824~$\mu$m (computationally expensive).  This ensures that any particles reflected from the right boundary region will be causally separated from the diagnostic region for the duration of the simulation (for a time $2\times 760\,\mu$m$/c\approx5$~ps).  The $p_x$-$x$ phasespace is shown in Fig.~\ref{fig:px-x-a} at 3.7~ps after the laser was incident on the critical interface, with the diagnostic region marked by dashed lines.  Note the large size of the plasma required compared to the diagnostic region location, along with the very hot return current reflecting off the right simulation boundary---even though a thermal particle boundary is being used.

We ran the simulations until 2~ps after the laser had finished hitting the plasma.  In all cases, hot particles were split into two after reaching a $\gamma$ of 1.4, 1.5, 1.6, 1.7, and 1.8 (i.e., very energetic particles were eventually split into 32 smaller particles); the splitting routine was executed every 10 time steps.  Contact the corresponding author for information about the source code and input files used for these simulations.

The particle acceleration mechanisms in these types of simulations are stochastic; therefore, we expect and indeed do observe large differences in particle statistics due to slightly different simulation configurations.  For example, we performed the causally separated simulation three different times with varied random number seeds and observed a factor of 2--4 variation in particle number in the tail of the momentum distribution over the diagnostic region.  For this reason we performed the simulations presented in this paper three times with different random number seeds.  Unless otherwise noted, visualizations presented here are of data averaged over three different runs; this averaging gives increased confidence that any observed deviations from the causally separated run are due to the particle boundary conditions.

\subsection{Effect of the absorber boundary condition}

To greatly reduce computation time and resources, we desire to shrink the simulation region shown in Fig.~\ref{fig:px-x-a}, but preserve the behavior from the causally separated run.  We truncate the plasma at a distance of 150~$\mu$m (29025 cells in $x$) and vary the length of the absorber, where each absorber is designed to stop all hot particles 5~$\mu$m short of the right boundary.  For all results shown here we use the linearly varying absorber from Sec.~\ref{sec:linear} and calculate the local temperature via Eq.~(\ref{Eq:lin-int}).  We quote the mean free path for each absorber, which as shown in Fig.~\ref{fig:f-and-h-lin} is 26\% of the entire absorber length.  We used an energy threshold of 6 times the local thermal velocity and re-emitted stopped particles at the local temperature.  Stopping was performed every time step for both electrons and ions
to give a large number of stopping loops for a fast particle traversing the absorbing region.  Particles are typically stopped every $\sim3$ time steps, but we perform a stopping loop every time step to more accurately assess the different methods.  Though it is much more important to use an absorber for electrons than for ions, we observed a sufficient number of hot ions reaching the thermal boundary to warrant stopping ions as well.  Stopping loops were delayed until hot particles approached the absorber region.  Particle recombination (for electrons) was executed every 5 time steps over the absorbing region; this dramatically reduces the simulation runtime as hot particles that have been split into 32 smaller particles are all stopped over a very short distance.

\begin{figure}
\includegraphics[width=\linewidth]{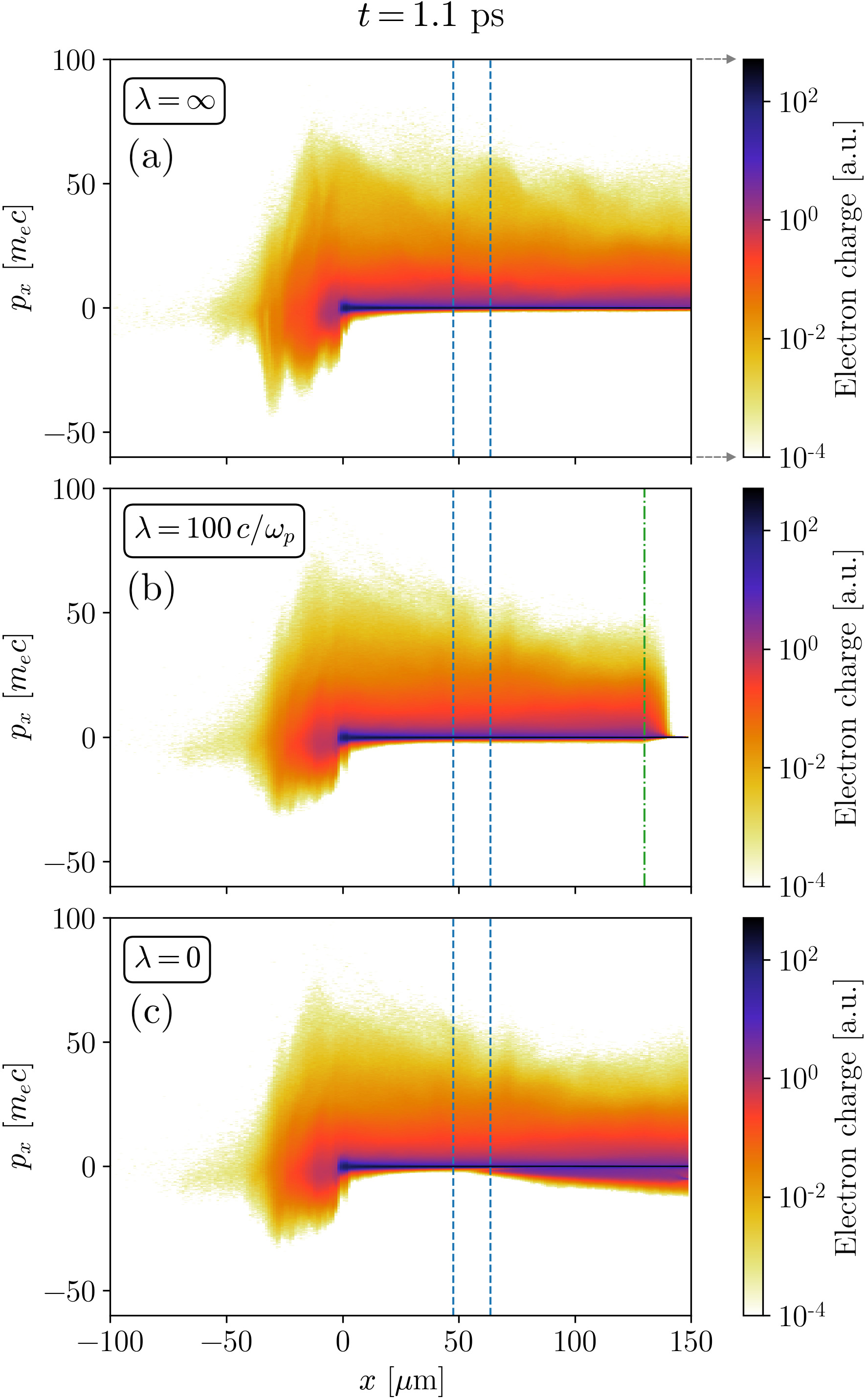}
\caption{\label{fig:px-x-1.1} The $p_x$-$x$ phasespace (single runs, not averaged) for the causally separated ($\lambda=\infty$), absorber, and no absorber ($\lambda=0$) simulations 1.1~ps after the incident laser.  A hot reflux of electrons is already shown to be entering the dashed diagnostic region for the truncated run with no absorber.}
\end{figure}

\begin{figure}
\includegraphics[width=\linewidth]{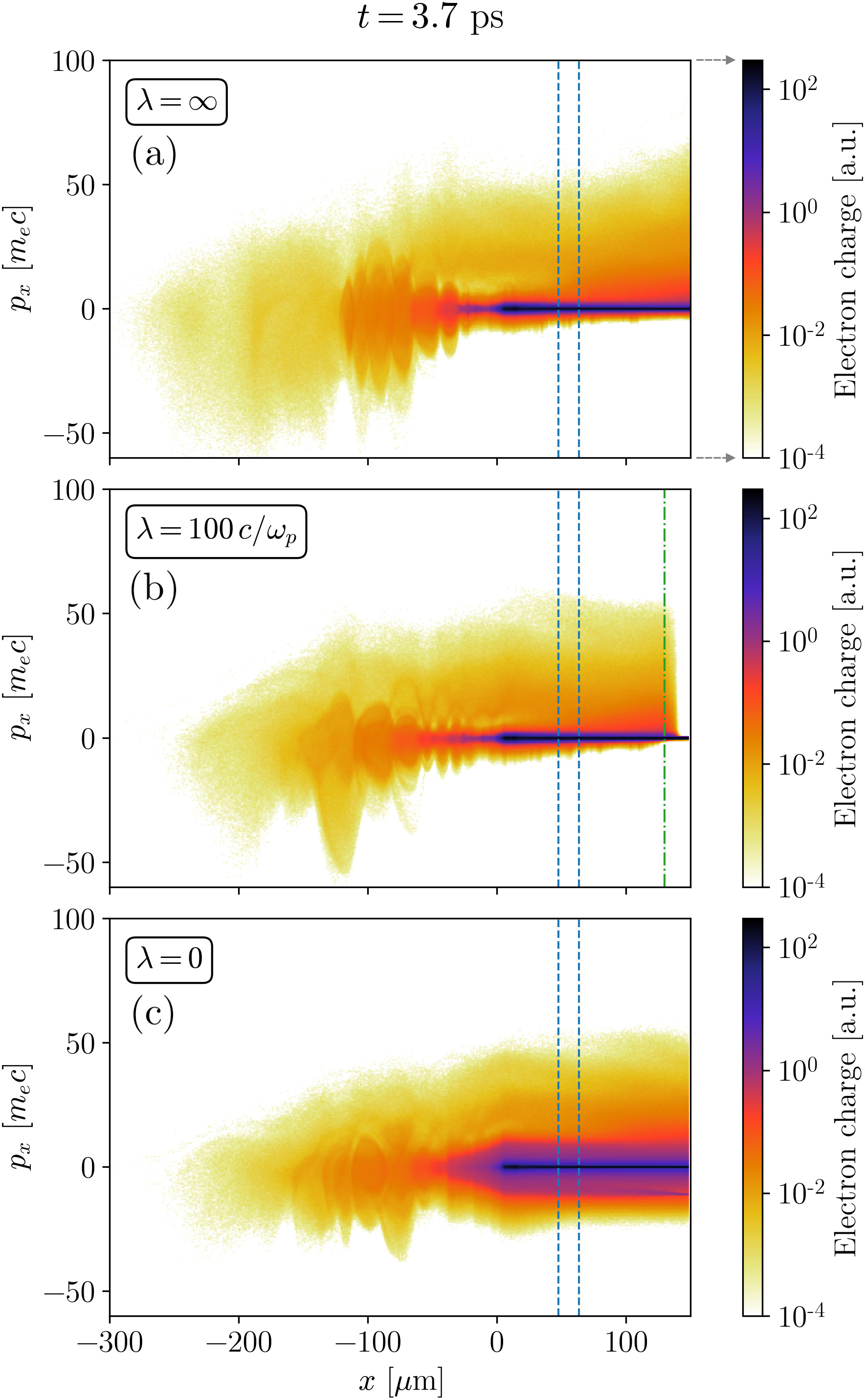}
\caption{\label{fig:px-x-3.7} The $p_x$-$x$ phasespace (single runs, not averaged) for the causally separated ($\lambda=\infty$), absorber, and no absorber ($\lambda=0$) simulations 3.7~ps after the incident laser.  The refluxing electrons for the truncated run have completely altered the particle phasespace; the returning hot electrons cyclically interact with the laser and re-enter the plasma, artificially heating the bulk plasma to a much higher temperature than in the casually separated or absorbing runs.}
\end{figure}

The $p_x$-$x$ phasespaces for the causally separated ($\lambda=\infty$, where we are zooming in on a particular region), absorber (with $\lambda=100\,c/\omega_p$) and no absorber/truncated ($\lambda=0$) simulations are shown in Figs.~\ref{fig:px-x-1.1} and \ref{fig:px-x-3.7} at two different times.  After just 1.1~ps, a hot reflux of electrons is visible in the truncated run [see Fig.~\ref{fig:px-x-1.1}(c)] that has already entered the diagnostic region.  These refluxing electrons are seen to completely overwhelm the simulation late in time [see Fig.~\ref{fig:px-x-3.7}(c)], while the simulation with the absorber [see Fig.~\ref{fig:px-x-3.7}(b)] is able to maintain an appropriate return current.  These plots are not averaged over three simulations, so sizeable variations within the pre-plasma are expected for the causally separated run due to differences in random number initialization with a different box size [note that the phasespace in the density upramp and surrounding region are identical in Figs.~\ref{fig:px-x-1.1}(b) and (c)].

\begin{figure}[htp]
\includegraphics[width=\linewidth]{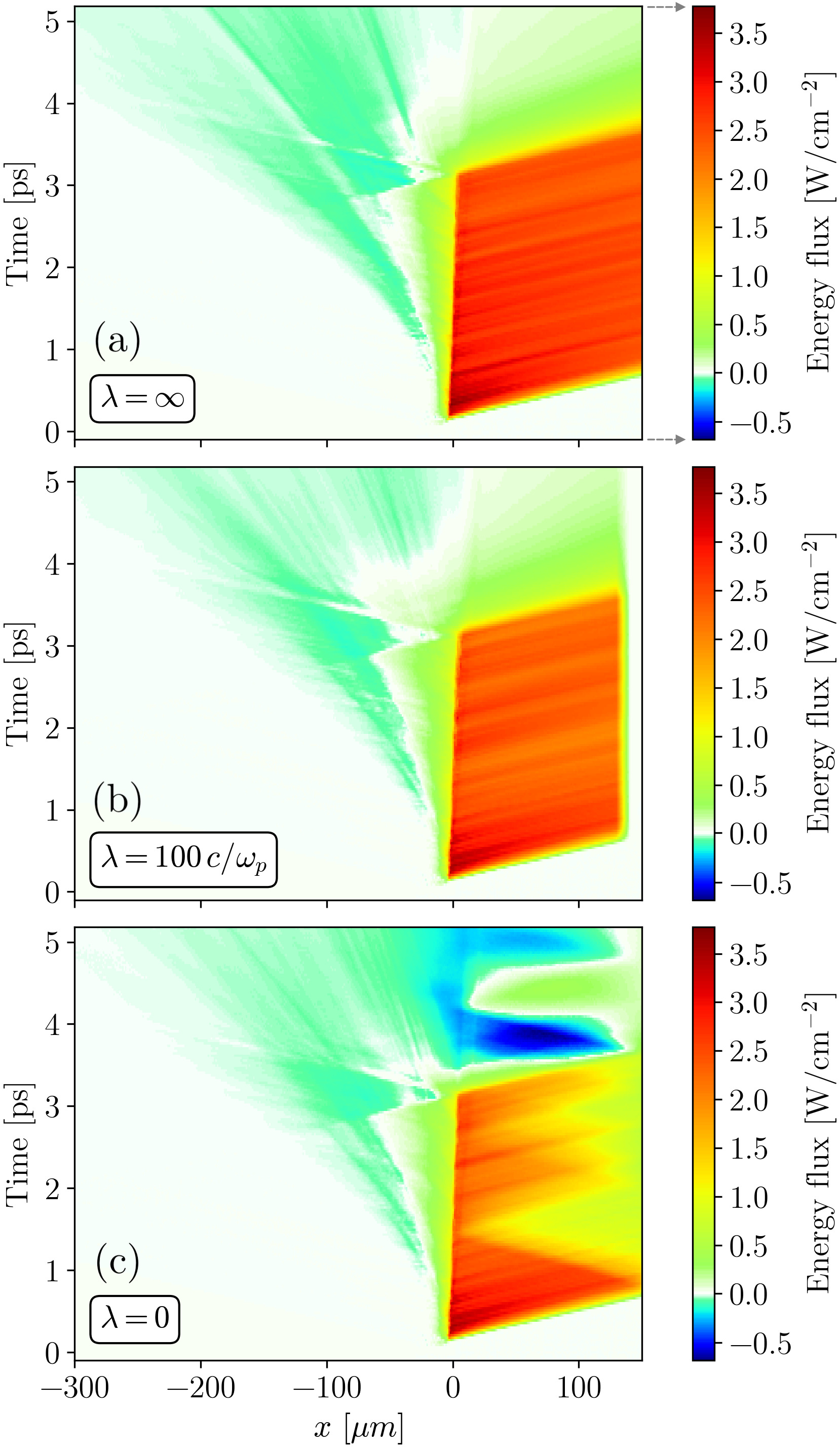}
\caption{\label{fig:s1-t} Forward particle energy flux as a function of position and time for three different cases.  For the truncated simulation ($\lambda=0$), the forward energy flux can be seen to be neutralized by a refluxing current emitted from the boundary.  The absorber effectively reduces the particle energy flux before the simulation boundary without a reflux current.}
\end{figure}

To better visualize temporal behavior, we plot the electron energy flux in the $x$ direction as a function of time and space for the causally separated, absorber, and no absorber simulations in Fig.~\ref{fig:s1-t}.  For the causally separated simulation, a steady stream of energy flux is observed to the right of the critical-density interface, which is slowly pushed forward in time.  Energetic electrons are also seen to escape to the left as the plasma expands.  This expansion is enhanced after the laser turns off.  When using the absorber with $\lambda=100\,c/\omega_p$, the energy flux looks qualitatively very similar to the causally separated run, except that the energy flux quickly decreases to zero in the absorber region.  In contrast, the truncated simulation ($\lambda=0$) shows that a large fraction of the forward energy flux is reflected from the right boundary (especially visible at 0.8~ps), so much so that it dramatically reduces the overall energy flux as it travels backward.  Once the first reflux arrives back to the laser-plasma interface at around 1.5~ps, the forward energy flux is then permanently altered.  This change in physics, as the hot return current interacts with and is accelerated by the laser, is the primary issue that the absorber is able to eliminate.  Finally, this hot reflux of electrons is also visible in the blue negative energy flux after the laser turns off in the truncated run.

\begin{figure}
\includegraphics[width=\linewidth]{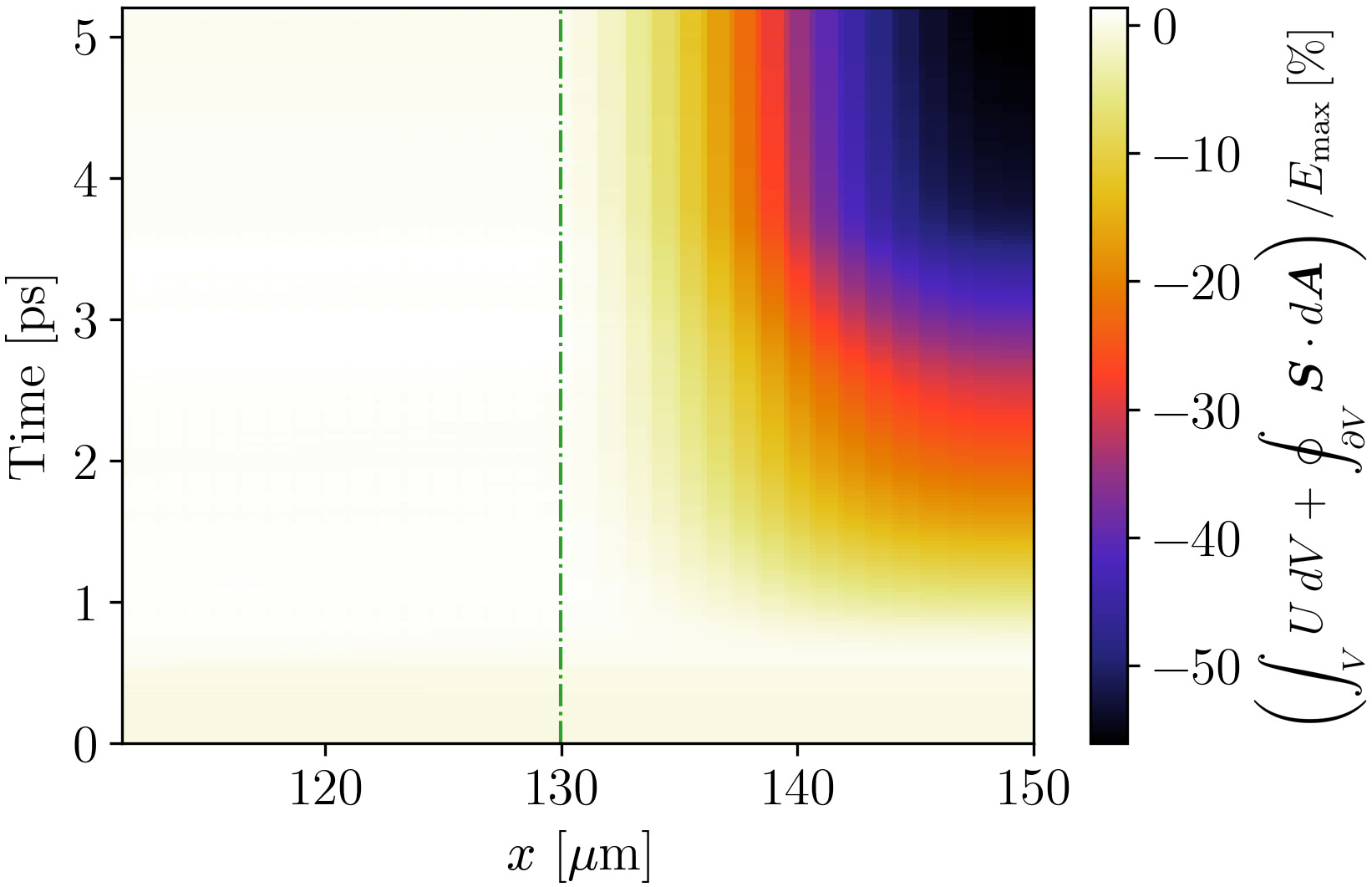}
\caption{\label{fig:energy} The scaled deviation in energy conservation [see Eq.~(\ref{Eq:energy})] as a function of time, including all points to the left of a given $x$ value (single run, not averaged).  To the left of the absorbing region, energy is well conserved ($<$1.4\% error), but in the absorbing region energy is steadily removed as particles are stopped.}
\end{figure}

We also examine energy conservation (fields plus particles) across the simulation region when the absorber boundary condition is in use.  To do this we compute the integral of energy density over a specific domain ($V$) and add the energy flux through the left and right boundaries of that domain ($\partial V$):
\begin{equation} \label{Eq:energy}
    \int_V U\,dV + \oint_{\partial V} \mathbf{S} \cdot d\mathbf{A},
\end{equation}
where $U$ is the energy density [$E^2/8\pi + B^2/8\pi + \sum (\gamma-1)m_e c^2$] and $\mathbf{S}$ is the energy flux [$\mathbf{E}\times \mathbf{B}/4\pi + \sum (\gamma-1)m_ec^2 \mathbf{p}/\gamma$].  We compute a running sum of this value over the simulation time (which should remain at zero) and then divide by the maximum energy present in the simulation box at any given time.  This gives a good measure of the energy conservation of the code, although it is not perfect since we only use data reported every 401 time steps (0.3~ps).  In Fig.~\ref{fig:energy} we plot Eq.~(\ref{Eq:energy}) as a function of time, where the right-hand side of volume $V$ (i.e., the location of $\partial V$ on the right) is given by the $x$ coordinate displayed for an absorber with mean free path $\lambda=100\,c/\omega_p$.  We can see that to the left of the absorber (dashed line), the deviation in the coarsely computed energy conservation is less than 1.4\%.  However, by including the absorber region we see that a large fraction of the energy is steadily removed as energetic particles are stopped.  Once again, it is this extended slowing of the particle beam that allows for an appropriate return current to develop, causing plasma to return back into the main simulation region.

\subsection{Variation of absorber parameters}

As mentioned in Sec.~\ref{sec:absorber} and Appendix~\ref{app:temp}, there are a variety of options for implementing the absorber region. 
When determining the energy threshold and re-emission temperature of the stopped particles, we can calculate the background temperature dynamically by weighting the distribution function with the proper velocity to some power,
or we can simply specify a constant value to use. Using a lower power (such as the fourth root) for the proper velocity will emphasize the bulk over a hot tail; more details are given in Appendix~\ref{app:temp}.  However, using the fourth-root temperature never improved the absorber performance for the simulations shown here, so we calculate the temperature in each cell as given by Eq.~(\ref{Eq:lin-int}).

We can also use the hazard function probability defined in Sec.~\ref{sec:hazard} or the linearly varying probability defined in Sec.~\ref{sec:linear} to stop the particles.  In our tests these two choices produce similar results, but overall the linearly varying absorber maintained the proper response for a longer time.  The main reason for this is that due to the periodicity in $y$, simulations using the hazard-function absorber exhibited a large and increasing transverse temperature in the absorbing region; the hazard-function absorber preferentially stops particles with large forward momentum, allowing energetic particles to stream transversely and for some accelerating/reflecting fields to develop (see last paragraph of Sec.~\ref{sec:hazard}).  For this reason we use the linearly varying absorber in this paper, which stops particles as a function of the magnitude of the velocity and not just the longitudinal component.

\begin{figure}
\includegraphics[width=\linewidth]{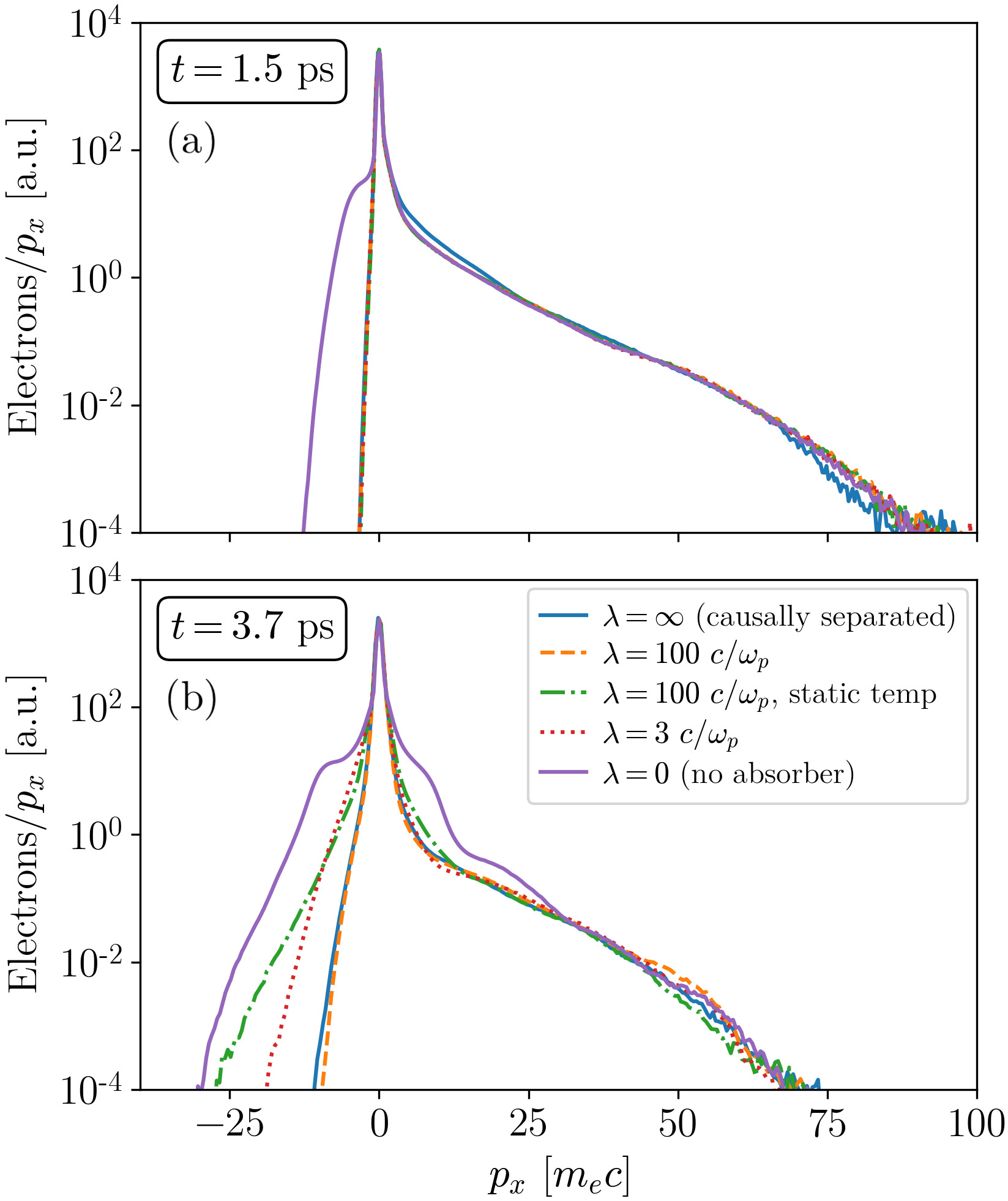}
\caption{\label{fig:variation} The $p_x$ phasespace for all electrons in the region 48--64~$\mu$m into the constant-density plasma for various schemes (a)~1.5~ps and (b)~3.7~ps after the laser was incident on the plasma.  Though the performance of all shown absorbers is nearly identical early in time, either using a static temperature threshold and re-emission or using a very short absorber gives improper results later in time.}
\end{figure}

We compare a combination of absorbers in Fig.~\ref{fig:variation}, where we show the $p_x$ momentum phasespace for all electrons in the diagnostic region at two different times.  Although all absorber schemes appear to perform equally well early in time, the return current is clearly hotter when constant values of the energy threshold and re-emission temperature are given.  For the static temperature simulation, we set the absorber to stop particles with energy greater than 0.6~keV and to re-emit particles at 0.1~keV (the original plasma temperature); in contrast the dynamic absorber stops particles moving at more than 6 times the locally computed thermal velocity.  Using a static temperature performs poorly because, as seen even in the absorbing region of Fig.~\ref{fig:px-x-3.7}(b), the plasma heats up significantly in response to the energetic electron beam.  Particles stopped and re-emitted at the original temperature are not moving fast enough to provide the necessary return current, and a nonphysical potential develops that accelerates electrons backward with too much energy.  Calculating the local temperature instead allows the absorber to accurately compensate for this dynamic behavior.

Although not shown here, we performed a series of simulations varying the mean free path of the absorber by factors of two between $\lambda=0.1\,c/\omega_p$ and $\lambda=200\,c/\omega_p$.  We observed that if the absorber had a mean free path $\lambda \gtrsim 6\,c/\omega_p$, it was able to closely match the causally separated momentum distribution when averaged over three separate runs.  However, individual simulations with $\lambda \lesssim 20\,c/\omega_p$ exhibited slightly greater variability in comparison to the causally separated data.  In our simulations, an absorber with a mean free path of $6\,c/\omega_p$ performed only $\sim30$ stopping events before nearly all particles were stopped, which was sufficient for a laser 3~ps in duration with $a_0=3$.  However, care must be taken for lasers of longer duration or higher intensity; Fig.~\ref{fig:variation} shows that some absorbers can perform well (a)~initially, but (b)~eventually fail due to the large amount of energetic particles striking the absorber.  Thus $\lambda \gtrsim \bigO(10\,c/\omega_p)$ gives a reasonable estimate of the appropriate mean free path, but the absorber length should be verified for each individual simulation.

\subsection{Best practices}

Here we make a few notes on best practices for performing simulations with the extended absorbing boundary condition.  We found it important to also causally separate the vacuum boundary (where the laser is injected) from the laser-plasma interface.  Even with absorbing particle boundary conditions at this vacuum boundary, most energetic particles that reached the vacuum boundary were immediately reflected back into the simulation space.  This is a combined effect of the laser potential at the wall and the electric field buildup from exiting particles (a nonnegligible number of particles are accelerated toward the laser from the pre-plasma region).  Refluxing from the vacuum boundary leads to a modified distribution at the laser-plasma interaction region, which then artificially inflates the forward electron energy flux in the target.


For simulations with a finite-width laser, absorber regions can also be placed at the transverse simulation edges to correctly handle the large flux of relativistic electrons expelled transversely from the laser spot.  However, the effectiveness of the absorber relies on having a large number of particles in each cell (for calculating the temperature).  If absorbers are placed at the transverse simulation boundaries, they may overlap with near-vacuum regions in and before the pre-plasma.
Thus for finite-size-target simulations with multiple absorbers, we found it is useful to transition the absorbers positioned along the transverse boundaries to stop and re-emit particles based on a static (rather than dynamically calculated) temperature in those near-vacuum regions.

Finally, the start of the absorber region should be located a reasonable distance away from where accurate plasma measurements are expected.  For example, when comparing Figs.~\ref{fig:px-x-3.7}(a) and \ref{fig:px-x-3.7}(b), the phasespace immediately in front of the absorbing region in (b) does not exactly mimic the causally separated phasespace in (a).  Examining the particle phasespace for irregularities near the absorber region can help determine the appropriate distance at which to measure plasma quantities.

\subsection{Future work}

The implementation described here, though effective, is by no means a comprehensive treatment or unique solution to the reflux problem.  Here we list some ideas that could be used to iterate on our proposed solution.  Particles could be re-emitted from a distribution that is hotter in the return direction than in the forward direction, assisting in establishing the appropriate return current.  Particles could be stopped preferentially based on their direction of motion.  We employed absorbers for both ions and electrons in these simulations, but the ion response and stopping could be explored in greater detail for long-time simulations.  Alternatives that are more computationally expensive could include applying a drag force to energetic particles over the length of the entire absorber or calculating the re-emission temperature from a position located before the absorber region.  Last, it may be possible to develop a thermal bath boundary where particles are re-emitted from a distribution determined from a region somewhere inside the plasma.

\section{Conclusion}

Particle-in-cell simulations are useful for investigating intense laser-plasma interactions in overdense plasmas, but a truncated plasma boundary can produce an unphysically hot return current.  This return current is present with absorbing, reflecting and thermal particle boundary conditions alike, and it can drastically alter simulation results.  We have devised an absorbing particle boundary condition that stops energetic particles over a defined region of the simulation space.  Stopping these particles over a sufficiently large distance allows the background plasma to generate a suitably cool return current that mimics the results of a semi-infinite, causally separated simulation.

Various different schemes were proposed for statistically selecting, stopping and re-emitting hot particles, with the best results given by the linearly varying absorber described in Sec.~\ref{sec:linear} that calculates the local temperature via Eq.~(\ref{Eq:lin-int}).  The appropriate mean free path of the absorber was explored, showing that an absorber with a mean free path of $\lambda \gtrsim \bigO(10\,c/\omega_p)$ gives proper results for our tests.
As simulation behavior can vary greatly depending on the application, care must be taken to ensure that the absorber parameters used for a particular case appropriately mimic the behavior of a semi-infinite boundary.

 \label{sec:conclusion}

\begin{acknowledgments}
The authors gratefully acknowledge guidance and feedback from A.~J.~Kemp and S.~C.~Wilks.

This work was performed in part under the auspices of the U.S. Department of Energy by Lawrence Livermore National Laboratory under Contract DE-AC52-07NA27344 and funded by the LLNL LDRD program with tracking code 19-SI-002 under Contract B635445. Additional support was given by DOE grant DE-SC0019010 and NSF grant 1806046.
\end{acknowledgments}

\section*{Data Availability Statement}

The data that support the findings of this study are available from the corresponding author upon reasonable request.

\appendix

\section{Computing local temperature} \label{app:temp}

In this appendix we describe how we calculate the local proper thermal velocity for use with the damping parameters.  First, we assume a Maxwellian distribution of the form
\begin{equation} \label{Eq:Max}
    f_0(u) = \frac{n_0}{\sqrt{2\pi \Bar{u}_0^2}} \exp \left( \frac{u^2}{2\Bar{u}_0^2} \right)
\end{equation}
for density $n_0$, proper velocity $u \equiv \gamma v$ and thermal velocity $\Bar{u}_0$.  We can approximate the average thermal velocity by summing $|u|$ over all particles in a cell.  For a Maxwellian of the same form as in Eq.~(\ref{Eq:Max}), we have that
\begin{equation} \label{Eq:lin-int}
    \int_{-\infty}^\infty |u| f_0(u) \, du = \sqrt{\frac{2}{\pi}}n_0\Bar{u}_0,
\end{equation}
which can be inverted to find $\Bar{u}_0$.

If instead the plasma is considered to be two distinct species with different thermal velocities and densities [e.g., including a beam with density $n_1$, thermal velocity $\Bar{u}_1$ and corresponding distribution $f_1(u)$], the above integral in Eq.~(\ref{Eq:lin-int}) could yield a distorted thermal velocity.  Another option is to perform the integral using the fourth-root of the thermal velocity, which gives
\begin{equation} \label{Eq:4th-int}
\begin{split}
    \int_{-\infty}^\infty |u|^{1/4} & \left[ f_0(u) + f_1(u) \right] \, du = \\
    & \frac{2^{1/8}\Gamma\left(\frac{5}{8}\right)}{\sqrt{\pi}} \left( n_0 \Bar{u}_0^{1/4} + n_1 \Bar{u}_1^{1/4} \right),
\end{split}
\end{equation}
where $\Gamma$ is the standard gamma function.  Since the densities add linearly but the thermal velocities add as the fourth-root, a high-energy beam should distort the sum less than in Eq.~(\ref{Eq:lin-int}).  Note that Eq.~(\ref{Eq:4th-int}) should be calculated for a single distribution and inverted to solve for $\Bar{u}_0$.  In practice, we found that the fourth-root calculation differed only slightly from the simple average of the thermal velocity due to the extremely low density of the beam.
However, the fourth-root calculation may be important for some parameter regimes.

\nocite{*}
\bibliography{references}

\end{document}